\documentclass[preprints,article,accept,moreauthors,pdftex]{./Definitions/mdpi} 

\firstpage{1} 
\pubvolume{1}
\issuenum{1}
\articlenumber{0}
\pubyear{2022}
\copyrightyear{2022}
\datereceived{} 
\dateaccepted{} 
\datepublished{} 
\hreflink{https://doi.org/} 
\pdfoutput=1

\Title{Theoretical study of dynamical and electronic properties of noncentrosymmetric superconductor NbReSi}

\TitleCitation{Theoretical study of dynamical and electronic properties of noncentrosymmetric superconductor NbReSi}

\Author{Surajit Basak $^{1}$\orcidA{} and Andrzej Ptok $^{1,\ast}$\orcidB{}}

\AuthorNames{Surajit Basak and Andrzej Ptok}

\AuthorCitation{Basak, S.; Ptok, A.}

\address{%
$^{1}$ \quad Institute of Nuclear Physics, Polish Academy of Sciences, W. E. Radzikowskiego 152, PL-31342 Krak\'{o}w, Poland; surajit.basak@ifj.edu.pl}

\corres{Correspondence: aptok@mmj.pl}

\abstract{
The noncentrosymmetric NbReSi superconductor with $T_{c} \simeq 6.5$~K is characterized by the relatively large upper critical magnetic field.
Its multigap features were observed experimentally.
Recent studies suggested realization of P$\bar{6}$2m or Ima2 symmetry.
We discuss the dynamical properties of both symmetries (e.g., phonon spectra).
In this paper, using the {\it ab initio} techniques, we clarify this ambiguity, and conclude that the Ima2 symmetry is unstable, and P$\bar{6}$2m should be realized.
The P$\bar{6}$2m symmetry is also stable in the presence of external hydrostatic pressure.
We show that NbReSi with the P$\bar{6}$2m symmetry should host phonon surface states for (100) and (110) surfaces.
Additionally, we discuss the main electronic properties of the system with the stable symmetry.
}


\graphicspath{{fig/}{./fig/}{.}}

\begin{document}


\section{Introduction}
\label{sec.intro}

Noncentrosymmetric superconductors are characterized by antisymmetric spin--orbit coupling (SOC)~\cite{yip.14,zhang.liu.22,smidman.salamon.17}, which gives rise to the topological superconducting pairing as a result of a mixture of spin-singlet and spin-triplet components~\cite{smidman.salamon.17,gorkov.rashba.01,ptok.kapcia.18}.
This behavior was first discovered in the heavy fermion compound CePt$_{3}$Si~\cite{bauer.hilscher.04}.
Recently, a large group of noncentrosymmetric superconductors have been discovered, e.g. 
Li$_{2}$(Pt$_{1-x}$Pd$_{x}$)$_{3}$B~\cite{togano.badica.04,yuan.agterberg.06},
BaPtSi$_{3}$~\cite{bauer.khan.09},
LaNiC$_{2}$~\cite{bonalde.ribeiro.11}, 
SrPtAs$_{2}$~\cite{nishikubo.kudo.11}, 
$R$PtSi ($R=$La, Ce, Pr, Nd, Sm, Gd)~\cite{klepp.parte.82,kneidinger.michor.13}, 
$A_{2}$Cr$_{3}$As$_{3}$ ($A=$K, Rb, Cs)~\cite{bao.liu.15,tang.bao.15,tang.bao.15b}, 
K$_{2}$Mo$_{3}$As$_{3}$~\cite{mu.ruan.18},
(Ta,Nb)Rh$_{2}$B$_{2}$~\cite{carnicom.xie.18},
Th$T$Si ($T=$Co, Ir, Ni, Pt)~\cite{domieracki.kaczorwski.16,domieracki.kaczorwski.18,ptok.domieracki.19}, or
CeRh$_{2}$As$_{2}$~\cite{khim.landaeta.21,ptok.kapcia.21}.

Recently many ternary noncentrosymmetric superconductors were discovered and studied.
Typically, the $MM'$Si class of materials (where $M$ and $M'$ are transition metals or rare earth metals) crystallise in several distinct structural symmetries, such as tetragonal PbClF-type (P4/nmm symmetry)~\cite{welter.venturini.93}, orthorhombic TiNiSi-type (Pnma symmetry)~\cite{morozkin.seropegin.99}, hexagonal ZrNiAl-type (P$\bar{6}$2m symmetry)~\cite{subbarao.wagner.85}, or orthorhombic TiFeSi-type (Ima2 symmetry)~\cite{subbarao.wagner.85}.
Among the mentioned symmetries, two are noncentrosymmetric (P$\bar{6}$2m and Ima2) and can give rise to unconventional triplet superconductivity.
These types of features were discussed in the case of Ta$T$Si ($T=$Re, Ru)~\cite{sajilesh.sinhg.21,sharma.sajilesh.22} or $T$RuSi ($T=$Ti, Nb, Hf, Ta)~\cite{shang.zhao.22}, both with Ima2 symmetry.

In our paper, we focus on the recently studied noncentrosymmetric NbReSi superconductor~\cite{shang.tay.22}, which exhibits superconducting properties below $T_{c} \simeq 6.5$~K~\cite{su.shang.21,shang.tay.22,sajilesh.motla.22}.
A relatively large upper critical magnetic filed was reported experimentally (around $12.5$~T~\cite{su.shang.21}, $13.5$~T~\cite{shang.tay.22}, or $8.1$~T~\cite{sajilesh.motla.22}).
The absence of spontaneous magnetic fields below T$_{c}$ from muon-spin relaxation ($\mu$SR) was observed~\cite{shang.tay.22}.
The superfuild density and the spin-lattice relaxation rate suggest nodeless superconductivity~\cite{shang.tay.22}.
The signatures of multigap superconductivity, evidenced by the field-dependent $\mu$SR rate and the electronic specific heat coefficient~\cite{shang.tay.22} can be related to the multiband Fermi level~\cite{su.shang.21}.

\begin{figure}[!t]
\centering
\includegraphics[width=\linewidth]{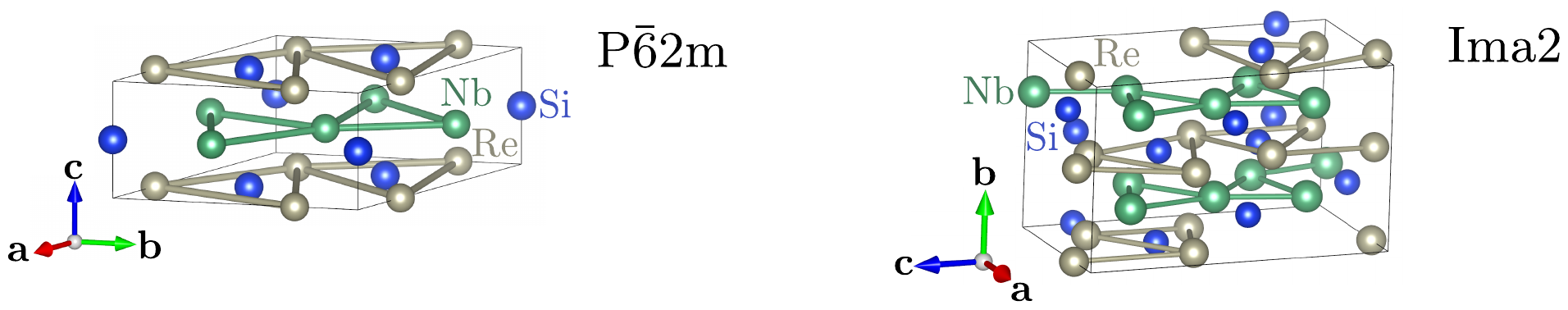}
\caption{
Schematic representation of NbReSi conventional cell with P$\bar{6}$2m and Ima2 symmetries (as labeled).
In both structures, the Re and Nb atoms form distorted kagome-like sublattices.
\label{fig.crys}
}
\end{figure}

However, the characterization of the studied samples suggest realization of the P$\bar{6}$2m symmetry (cf. Ref.~\cite{su.shang.21,shang.tay.22}) or Ima2 (cf. Ref.~\cite{sajilesh.motla.22}) -- structures presented in Fig.~\ref{fig.crys}.
We clarify this ambiguity using the {\it ab initio} techniques.
From the study of dynamical properties, we claim that the P$\bar{6}$2m symmetry is stable in this system.
We also discuss the dynamical and electronic properties of the system with stable symmetry.

The paper is organized as follows.
Details of the techniques used are provided in Sec~\ref{sec.method}.
Next, in Sec.~\ref{sec.res} we present and discuss our theoretical results.
Finally, a summary is provided in Sect.~\ref{sec.sum}.


\section{Calculation details}
\label{sec.method}

First-principles (DFT) calculations are performed using the projector augmented-wave (PAW) potentials~\cite{blochl.94} implemented in the Vienna Ab initio Simulation Package ({\sc Vasp}) code~\cite{kresse.hafner.94,kresse.furthmuller.96,kresse.joubert.99}.
Calculations are made within the generalized gradient approximation (GGA) in the Perdew, Burke, and Ernzerhof (PBE) parameterization~\cite{pardew.burke.96}.
The calculations, including SOC, were performed with the energy cut-off set to $600$~eV.

Initially, the crystal structure and atom positions were optimised. 
In the case of the P$\bar{6}$2m symmetry, the primitive cell containing three formula units was optimized, with the $6 \times 6 \times 12$ {\bf k}--point grid in the Monkhorst--Pack scheme~\cite{monkhorst.pack.76}.
Similarly, for the Ima2 symmetry, we used a conventional unit cell with $6 \times 6 \times 3$ {\bf k}--point grid.
As the convergence condition of an optimization loop, we take the energy difference 
of $10^{-5}$~eV and $10^{-7}$~eV for ionic and electronic degrees of freedom, respectively.
Optimized structure parameters are collected in Sec.~\ref{sec.res_crys}

The interatomic force constants (IFC) are calculated within the Parlinski-Li-Kawazoe method~\cite{parlinski.li.97} implemented in {\sc Phonopy} package~\cite{togo.tanaka.15}.
Force constants were obtained from first-principles calculations of the Hellmann--Feynman forces by {\sc VASP} and used to build a dynamical matrix of the crystal. 
Phonon frequencies were obtained by diagonalization of the dynamical matrix.
Calculations were performed using the supercell technique. 
In the case of the P$\bar{6}$2m symmetry, the supercell based on $2 \times 2 \times 3$ primitive cells was used.
For the Ima2 symmetry, we used a supercell based on $2 \times 1 \times 2$ conventional cells.
In both cases, the reduced $3 \times 3 \times 3$ {\bf k}-grid was used.
Furthermore, dynamical properties were evaluated withing {\sc Alamode} software~\cite{tadano.gohda.14}, using the multidisplacement method.
Finally, to study the surface states of phonons, the surface Green's function for a semi-infinite system~\cite{sancho.sancho.85} was calculated using {\sc WannierTools}~\cite{wu.zhang.18}.


\section{Results and discussion}
\label{sec.res}

\subsection{Crystal structure}
\label{sec.res_crys}

After optimizing the crystal structures, we found:
\begin{itemize}
\item for the P$\bar{6}$2m symmetry (space group No. 189): $a = b = 6.872$~\AA, and $c = 3.310$~\AA, while experimental values are 
$a = b = 6.719$~\AA, and $c = 3.485$~\AA~\cite{su.shang.21};
Nb atoms are in Wyckoff positions $3g$: ($0.4020$,$0$,$1/2$), Re atoms in Wyckoff positions $3f$: ($0.7411$,$0$,$0$), while Si atoms in the two non-equivalent Wyckoff positions $2c$: ($1/3$,$2/3$,$0$) and $1b$: ($0$,$0$,$1/2$).
\item for the Ima2 symmetry (space group No. 46): $a = 6.990$~\AA, $b = 11.618$~\AA, $c = 6.726$~\AA, while the reported values are $a = 6.925$~\AA, $b = 11.671$~\AA, and $c = 6.694$~\AA~\cite{sajilesh.motla.22};
Nb atoms are in three non-equivalent Wyckoff positions $4b$: ($1/4$,$0.1959$,$0.7093$), $4b$: ($1/4$,$0.7873$,$0.7130$) and $4a$: ($1/4$,$-0.0033$,$0.0919$), Re atoms in the two non-equivalent Wyckoff positions $4a$ ($0$,$0$,$0.7547$) and $8c$: ($0.5344$,$0.8732$,$0.3776$), while Si atoms in two non-equivalent positions $4b$: ($1/4$,$-0.0259$,$0.4920$) and $8c$: ($0.0037$,$0.1677$,$0.0064$).
\end{itemize}
In the case of the P$\bar{6}$2m symmetry, the primitive unit cell is equivalent to the conventional cell and contains three formula units.
Contrary to this, for the Ima2 symmetry, the primitive cell contains six formula units, while the conventional one is twice bigger.
The obtained crystal parameters are used in the next part of the paper as reference ones.


\begin{figure}[!b]
\centering
\includegraphics[width=\linewidth]{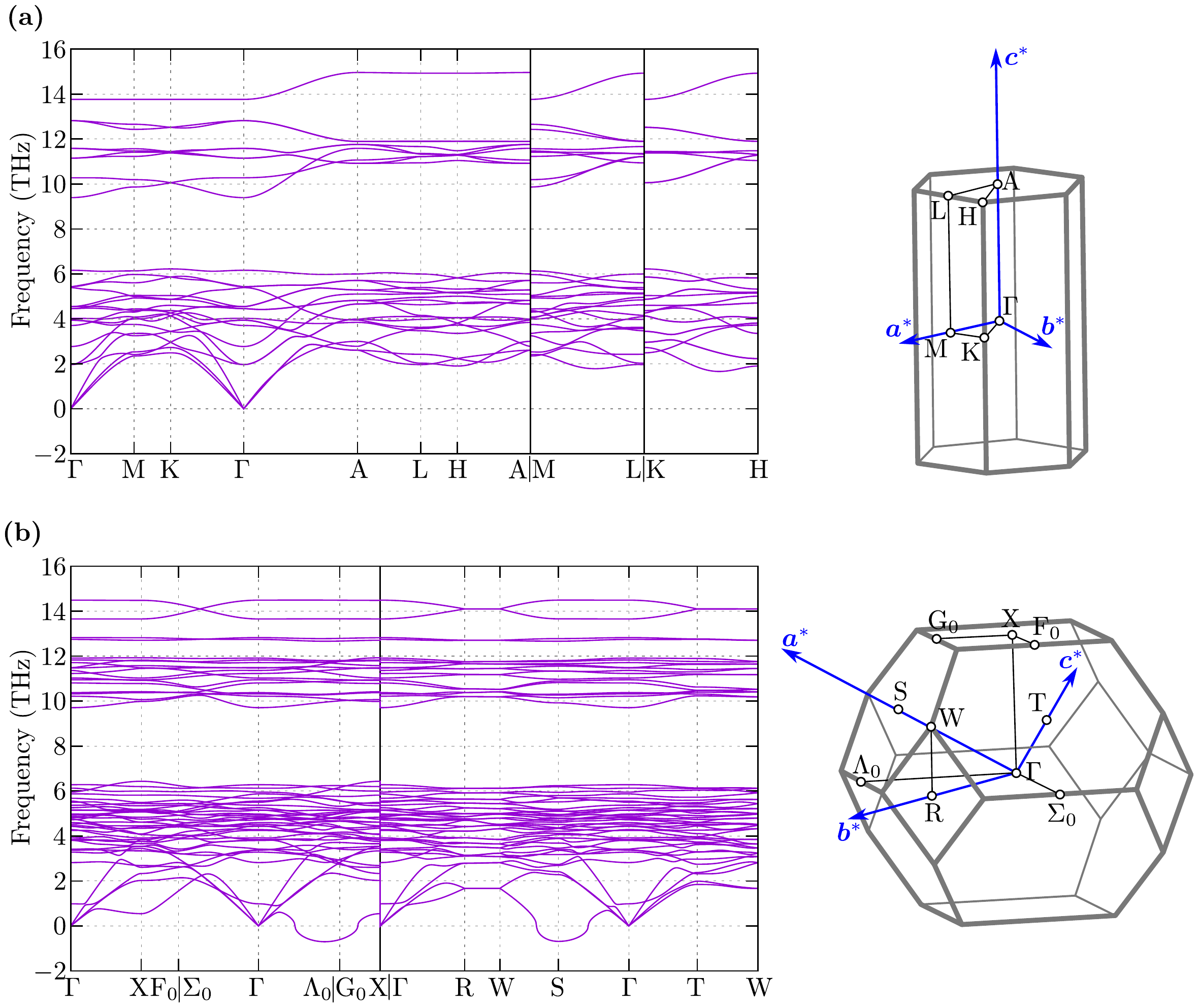}
\caption{
The phonon dispersion along high symmetry directions (left panels) and the Brillouin zone (right panels) for P$\bar{6}$2m and Ima2 (top and bottom panels, respectively).
\label{fig.ph_band}
}
\end{figure}

\subsection{Dynamical properties}

The phonon dispersions for both symmetries are presented in Fig.~\ref{fig.ph_band}.
Regardless of the symmetry, the phonon branches are collected into several groups separated by gaps.
In the case of P$\bar{6}$2m symmetry, acoustic modes exhibit a linear behavior around the $\Gamma$ point.
For the Ima2 symmetry, one of the acoustic modes along the $\Gamma$--$\Lambda_{0}$ path (continuing along the G$_{0}$-X path) poses the soft mode.
However, this soft mode is also visible around the S-point. 
From the dynamical point of view, NbReSi is unstable with the Ima2 symmetry.

In both symmetries, Re and Nb atoms form distorted kagome-like sublattices (see Fig.~\ref{fig.crys}), and should exist a relation between these two structures. 
In fact, there is a group-subgroup relationship between the discussed symmetries~\cite{ivantchev.koumova.00}, which is, P$\bar{6}$2m$\rightarrow$Amm2$\rightarrow$Ima2.
This allows Ima2 to emerge from P$\bar{6}$2m as a consequence of atom displacements.
Nevertheless, P$\bar{6}$2m is stable and does not exhibit any soft modes that can lead to Amm2 or Ima2 symmetries.

\begin{figure}[!t]
\centering
\includegraphics[width=0.6\linewidth]{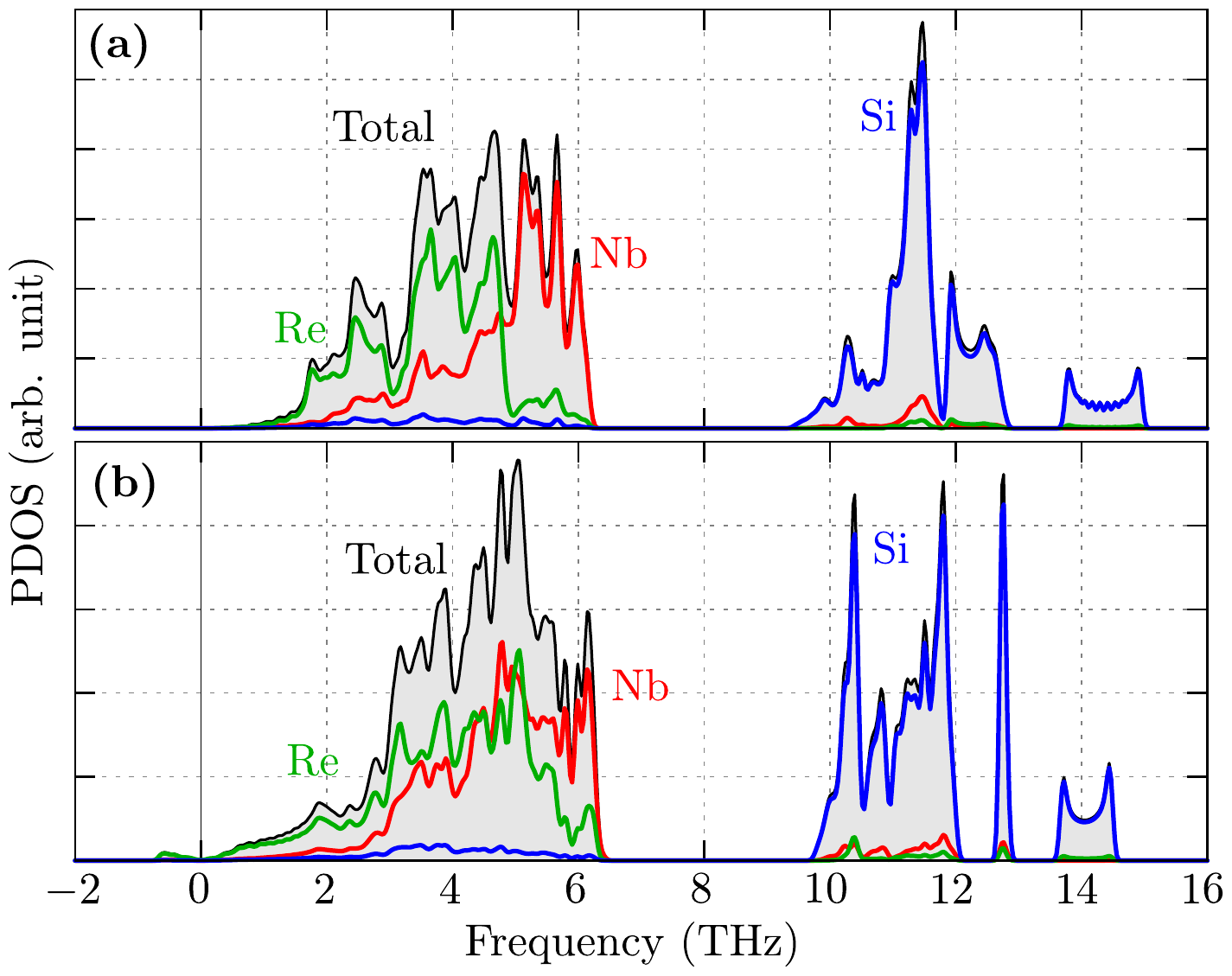}
\caption{
The phonon density of states (PDOS) for P$\bar{6}$2m and Ima2 (top and bottom panels, respectively).
\label{fig.ph_dos}
}
\end{figure}

The above mentioned properties are also reflected in the phonon density of states (PDOS), presented in Fig.~\ref{fig.ph_dos}.
A more precise analysis uncovers the contribution of separate atoms in vibration modes.
The lower-frequency modes are realized by Re and Nb atoms (relatively heavy atoms).
As can be expected, high-frequency modes are realized by lighter atoms, i.e. Si.
The PDOS are qualitatively comparable for both symmetries.
As we can see, the soft mode in the Ima2 structure is realized by vibrations of Nb atoms.

The irreducible representations of the phonon modes at the $\Gamma$ point are given as~\cite{kroumova.aroyo.03}:
\begin{itemize}
\item for the P$\bar{6}$2m symmetry:
\begin{eqnarray}
\label{eq.modes_p62m} \Gamma_\text{acoustic} &=& A''_\text{2} + E' \\
\nonumber \Gamma_\text{optic} &=& 2 A'_\text{1} + 2 A'_\text{2} + A''_\text{1} + 3 A''_\text{2} + 6 E' + 2 E'' ,
\end{eqnarray}
where $A''_\text{2} + E'$ modes are infra-red active, while $E' + E''$ modes are Raman active.
\item for the Ima2 symmetry:
\begin{eqnarray}
\label{eq.modes_ima2} \Gamma_\text{acoustic} &=& A_\text{1} + B_\text{1} + B_\text{2} \\
\nonumber \Gamma_\text{optic} &=& 14 A_\text{1} + 11 A_\text{2} + 11 B_\text{1} + 15 B_\text{2} ,
\end{eqnarray}
where $A_\text{1} + B_\text{1} + B_\text{2}$ modes are infra-red active, while $A_\text{1} + A_\text{2} + B_\text{1} + B_\text{2}$ modes are Raman active.
\end{itemize}
For lower symmetry (i.e. Ima2), in practice all modes are Raman active.
Here we should note that in both symmetries primitive cells contain different numbers of atoms.
As a consequence, the Raman spectra can be used as a tool to confirm the symmetry realized by NbReSi.

\begin{figure}[!t]
\begin{adjustwidth}{-\extralength}{0cm}
\centering
\includegraphics[width=\linewidth]{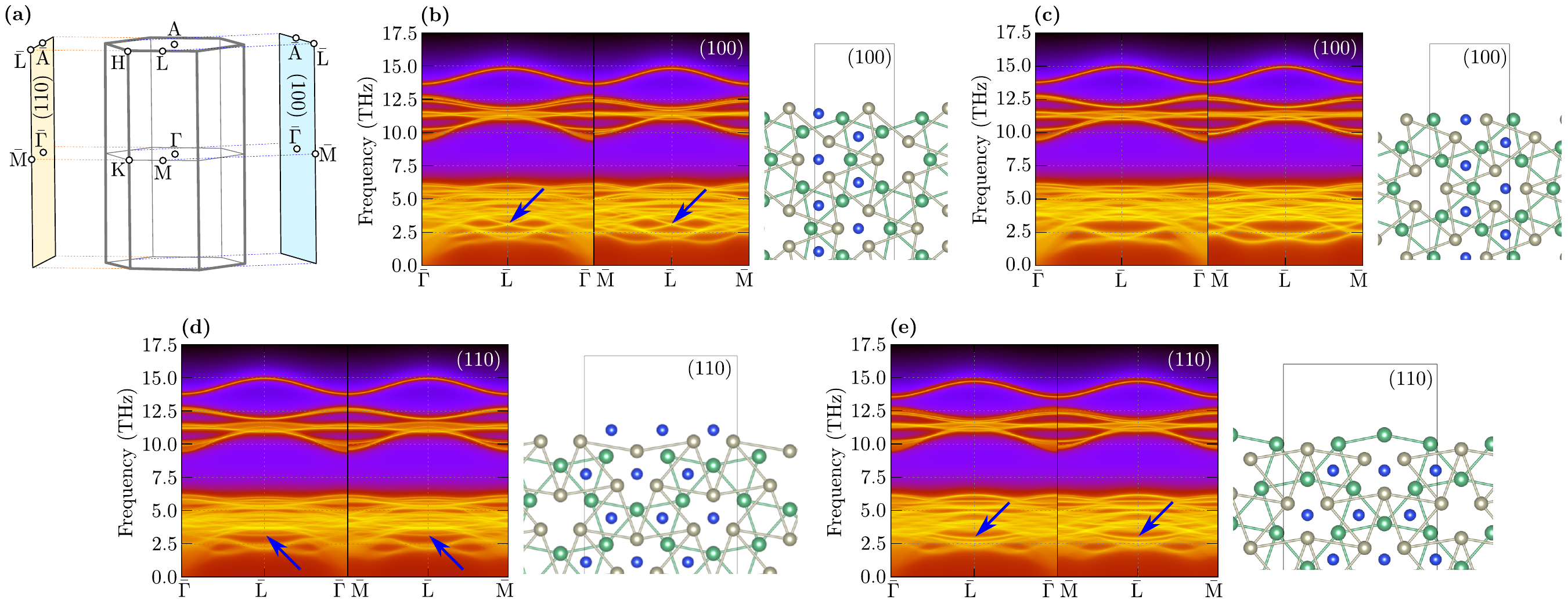}
\end{adjustwidth}
\caption{
Projection of the 3D bulk Brillouin zone on the 2D surface Brillouin zones for (100) and (110) surfaces (a).
Panels (b)-(e) present surface Green functions for different surfaces (as labeled) and corresponding terminations (right schemats).
Blue arrows show phonon surface states at $\bar{\text{L}}$ point.
}
\label{fig.ph_ss}
\end{figure}

\subsubsection*{\it Phonon surface states}

Realization of hexagonal symmetry by NbReSi can give rise to phonon surface states~\cite{li.wang.20}.
The calculated surface Green functions are presented in Fig.~\ref{fig.ph_ss}.
The phonon dispersion (Fig.~\ref{fig.ph_band}) is very complex and concentrated in the low frequency range (below $7.5$~ THz).
However, for surfaces (100) and (110), the phonon edge states are clearly visible at several places.
For example, at the $\bar{\text{L}}$ point, where the phonon surface states are realized by separated modes with relatively large intensity (marked by blue arrows in Fig.~\ref{fig.ph_ss}).
Interestingly, for some terminations, this mode is not realized, see Fig.~\ref{fig.ph_ss}(c).

The relatively small frequencies of the mentioned phonon surface modes suggest their strong connection with Nb or Re atoms on the surface.
For the surface (100), where the bulk M and $\Gamma$ points are projected on the surface $\bar{\Gamma}$ point, we observed a Dirac-like structure (Fig.~\ref{fig.ph_ss}(b)).
A more precise analysis of the surface band structure uncovers avoided crossing of two edge states.
Contrary to this, for the (110) surface, where bulk K and M points are projected on the surface $\bar{\text{M}}$ point, while bulk K and $\Gamma$ points are projected on surface $\bar{\Gamma}$ point, we observed only one separated surface state (Fig.~\ref{fig.ph_ss}(c) and Fig.~\ref{fig.ph_ss}(d)).

\subsubsection*{\it Role of hydrostatic pressure}

\begin{figure}[!h]
\centering
\includegraphics[width=\linewidth]{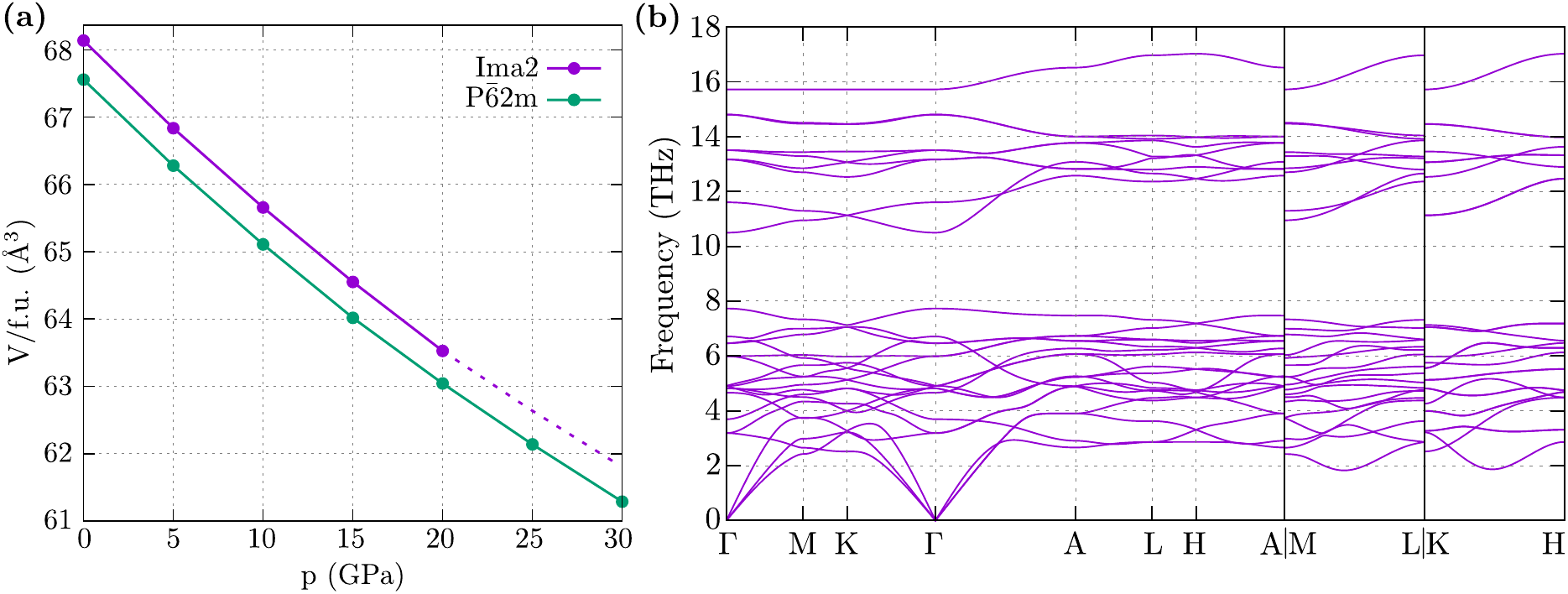}
\caption{
Influence of external hydrostatic pressure on NbReSi.
(a) Pressure dependence of the unit cell volume (per formula unit) for system with P$\bar{6}$2m and Ima2 symmetries (as labeled).
(b) Phonon dispersion curves for system with P$\bar{6}$2m under pressure 30~GPa.
}
\label{fig.press}
\end{figure}

Now we briefly discuss the impact of external hydrostatic pressure on NbReSi.
Under external pressure, the volume of the system decreases monotonically (Fig.~\ref{fig.press}(a)).
In the absence of pressure, the energy for both symmetries is comparable.
However, comparison of their volumes (per formula unit, see Fig.~\ref{fig.press}(a)) clearly show that the unit cell of P$\bar{6}$2m is always more dense. 
This feature has an important impact under external pressure and indicates a smaller enthalpy (i.e., the sum of the ground states energy and $pV$ terms) of the system with P$\bar{6}$2m than Ima2.
Furthermore, above some pressure (around 20~GPa), the system with the Ima2 symmetry was impossible to optimize, and the structure goes to the P$\bar{6}$2m symmetry.
This suggests instability of the NbReSi system with the Ima2 symmetry even under pressure.

Fig.~\ref{fig.press}(b) shows the dispersion cerves for the P$\bar{6}$2m symmetry under pressure 30~GPa, which are comparable with the one obtained in the ambient pressure (i.e. Fig.~\ref{fig.ph_band}(a)).
The range of realized phonon frequencies increase under pressure, as a result of the decreased volume of the system.
Nevertheless, the system is stable and does not exhibit phonon softening.

\begin{figure}[!h]
\centering
\includegraphics[width=0.6\linewidth]{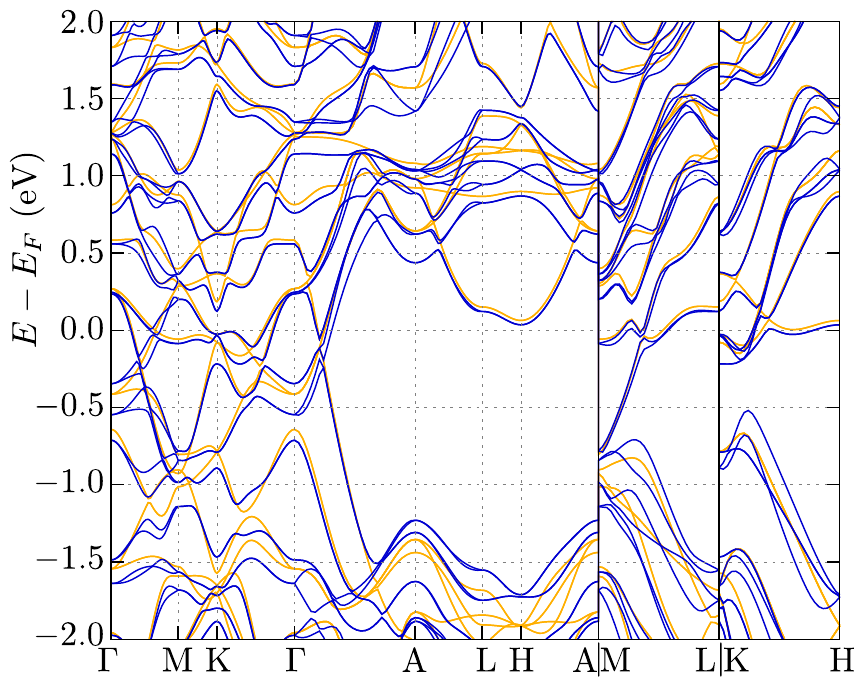}
\caption{
The electronic band structure along high symmetry directions for the P$\bar{6}$2m symmetry in the absence and presence of spin--orbit coupling (orange and blue lines, respectively).
\label{fig.el_band}
}
\end{figure}

\subsection{Electronic properties}

Now, we briefly describe the electronic properties of the NbReSi system with stable the P$\bar{6}$2m symmetry.
The obtained electronic band structure (Fig.~\ref{fig.el_band}) is in agreement with the initial one presented in Ref.~\cite{su.shang.21}.

The SOC band splitting near the Fermi level is estimated to be $180$~meV.
This is a relatively high value of the SOC in comparison to the other noncentrosymmetric superconductors~\cite{smidman.salamon.17}, suggesting the realization of topological superconductivity in NbReSi.
A significant role of the antisymmetric SOC is well visible along the $c$ direction (e.g., the M--L or K--H paths in Fig.~\ref{fig.el_band}).
A large SOC is observed in the $k_{z} =0$ plane (e.g. along the $\Gamma$--M--K--$\Gamma$ path), in contrast to a relatively suppressed value of the SOC for $k_{z} = \pi / c$ (e.g. along the A--L--H--A).

\begin{figure}[!b]
\centering
\includegraphics[width=0.7\linewidth]{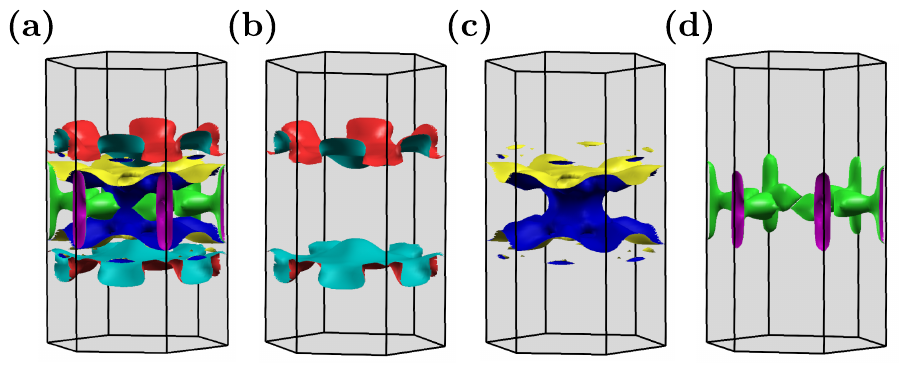}
\caption{
The Fermi surface for the P$\bar{6}$2m symmetry (in the absence of spin--orbit coupling).
Panel (a) presents the total Fermi surface, while panels (b)-(d) show separated pockets.
\label{fig.el_fermi}
}
\end{figure}

The Fermi surface is presented in Fig.~\ref{fig.el_fermi}.
In the absence of SOC, the Fermi surface is composed of three pockets. 
Introduction of the SOC leads to band decoupling, whereas the Fermi surface preserves the shape unchanged qualitatively.
Surprisingly, the Fermi surface exhibits quasi-one-dimensional features (i.e., the Fermi surface pocket are given by relatively flat pieces for $k_{z} = \text{const.}$, see Fig.~\ref{fig.el_fermi}(b)).

The features of the electronic band structure, as well as the Fermi surface topology, make NbReSi very similar to the $A_{2}$Cr$_{3}$As$_{3}$ ($A=$K, Rb, Cs)~\cite{jiang.cao.15,cuono.autieri.19,xu.wu.20,taddei.binghua.22} and K$_{2}$Mo$_{3}$As$_{3}$~\cite{yang.feng.19,taddei.binghua.22} compounds.
NbReSi does not exhibit any magnetic order features (the ground state is nonmagnetic), while mentioned $A_{2}$Cr$_{3}$As$_{3}$ poses magnetic order~\cite{wu.le.15}.
Regardless of this, the electronic band structure at $k_{z} = \pi/c$ is characterized by a relatively large gap, which was observed in both cases. 
For NbReSi, the states around the Fermi level are mostly composed of Re-$5d$ and Nb-$4d$ orbitals~\cite{su.shang.21}, while for $A_{2}$Cr$_{3}$As$_{3}$, Cr-$3d$ orbitals have the greatest contribution~\cite{jiang.cao.15}.
The Fermi surface of NbReSi is very similar to $A_{2}$Cr$_{3}$As$_{3}$~\cite{jiang.cao.15,cuono.autieri.19,xu.wu.20,taddei.binghua.22} or K$_{2}$Mo$_{3}$As$_{3}$~\cite{yang.feng.19,taddei.binghua.22} (both cases with P$\bar{6}$2m symmetry).

Here we should mention that despite having the same symmetry, the two structures possess different intrinsic features: e.g., $A_{2}$Cr$_{3}$As$_{3}$ contains quasi-one-dimensional chain of Cr atoms along the $c$ direction, while NbReSi has layers of distorted kagome-like sublattice of Nb and Re atoms in the $ab$ plane (see Fig.~\ref{fig.crys}).
This causes detailed differences in the electronic band structure.
Quasi-one-dimensional chains in $A_{2}$Cr$_{3}$As$_{3}$ are related to the nearly flat bands within the $ab$ plane and a strong $k_{z}$ dependence of the electron dispersion.
In NbReSi, too, we observe strong $k_{z}$-dependence of the electron dispersion.
Additionally, the absence of characteristic band structure features for the kagome-like structure is observed (i.e. absence of the flat bands). 
This can be related to the relatively large distances between atoms in the distorted kagome-like planes ($3.63$~\AA\ and $4.47$~\AA\ for Nb--Nb and Re--Re pairs, respectively; cf. Fig.~\ref{fig.crys}).

The smaller distance between the atoms is related to the bonding between Re and Si (distance around $2.43$~\AA), and can have a relatively strong impact on the electronic properties.
Indeed, the charge density distribution analyses (not presented) provide signature of a strong bonding within these pairs.
Such a structure with a strong bonding between the atoms along $c$ can be responsible for the quasi-one-dimensional character of NbReSi visible on the Fermi surface.


\section{Summary and conclusions}
\label{sec.sum}

We discussed the basic properties of the recently experimentally studied noncentrosymmetric superconductor NbReSi.
The experiments suggest realization of the P$\bar{6}$2m symmetry (cf. Ref.~\cite{su.shang.21,shang.tay.22}) or Ima2 (cf. Ref.~\cite{sajilesh.motla.22}) symmetry.
Using the {\it ab initio} technique, we show that NbReSi is stable with the P$\bar{6}$2m symmetry, while the Ima2 phase exhibits (imaginary) phonon soft modes.
This can be verified in a relatively simple way by the Raman scattering measurements.
The NbReSi with the P$\bar{6}$2m symmetry is stable also under hydrostatic pressure.
We also found that the phonon surface states can be realized by NbReSi with the P$\bar{6}$2m symmetry, for (100) and (110) surfaces.

NbReSi with stable the P$\bar{6}$2m symmetry exhibits the electronic band structure and the Fermi surface very similar to quasi-one-dimensional $A_{2}$Cr$_{3}$As$_{3}$~\cite{wu.yang.15}.
Surprisingly, the Fermi surface of NbReSi uncovers quasi-one-dimensional features, which can be associated with the realization of quasi-one-dimensional chains of Re--Si, with a strong bonding between the atoms.
Additionally, the relatively large value of spin--orbit coupling, as well as similarities to $A_{2}$Cr$_{3}$As$_{3}$ promote this compound as a good candidate for the realization of unconventional superconductivity.

\vspace{6pt} 



\authorcontributions{
A.P. initialized this project; S.B. and A.P. realized theoretical calculations; A.P. prepared the first version of the manuscript. All authors have read and agreed to the published version of the manuscript.
}

\funding{
This work was supported by National Science Centre (NCN, Poland) under Projects No.
2021/43/B/ST3/02166 (A.P.). 
}

\institutionalreview{Not applicable}

\informedconsent{Not applicable}

\dataavailability{Not applicable} 

\acknowledgments{
Some figures in this work were rendered using {\sc Vesta}~\cite{momma.izumi.11}.
A.P. appreciates funding in the framework of scholarships of the Minister of Science and Higher Education (Poland) for outstanding young scientists (2019 edition, No. 818/STYP/14/2019).
}

\conflictsofinterest{The authors declare no conflict of interest.} 


\begin{adjustwidth}{-\extralength}{0cm}

\reftitle{References}


\bibliography{biblio.bib}

\end{adjustwidth}
\end{document}